\documentclass[fleqn,usenatbib]{mnras}

\usepackage{newtxtext,newtxmath}
\usepackage{xcolor}
\usepackage[T1]{fontenc}
\usepackage{ae,aecompl}
\usepackage{hyperref}
\usepackage{threeparttable}

\newcommand{\EQ}[1] {equation~(\ref{#1})}
\newcommand{\SEC}[1] {Section~\ref{#1}}
\newcommand{\APP}[1] {Appendix~\ref{#1}}
\newcommand{\FIG}[1] {Figure~\ref{#1}}
\newcommand{\TAB}[1] {Table~\ref{#1}}

\definecolor{dkblue}{RGB}{54,169, 86}

\usepackage{natbib}
\bibpunct{(}{)}{;}{a}{}{,}

\usepackage{graphicx}	
\usepackage{dblfloatfix}             
\usepackage{float}
\usepackage{amsmath}	
\usepackage{wrapfig}

\title[FRB detection in the presence of coloured noise]{Fast radio burst 
	detection in the presence of coloured noise}

\author[Cfzhang et al.]{
	C.~F. Zhang$^{1,2}$,
	J.W. Xu$^{1,2}$,
	Y.~P. Men$^{3, 1,2}$,
	X.~H. Deng$^{4}$,
	Heng Xu$^{1,2}$,
	J.~C.Jiang$^{1,2}$,\newauthor
	B.~J.Wang$^{1,2}$,
	K.~J. Lee$^{1}$\thanks{E-mail: kjlee@pku.edu.cn},
	J.Li$^{5}$,
	J.~P. Yuan$^{5}$,
	Z.~Y. Liu$^{5}$,
	Y.~X.Huang$^{6}$,
	Y.~H.Xu$^{6}$,
	Z.~X.Li$^{6}$,\newauthor
	L.~F.Hao$^{6}$,
	J.~T. Luo$^{7}$, 
	S.Dai$^{7}$,
	R. Luo$^{8}$,
	Hassan Zakie$^9$ ,
	Z.~Y. Ma $^{4}$
	\\
	$^{1}$Department of Astronomy, Peking University, Beijing 100871, P.R. China 
	\\
	$^{2}$Kavli Institute for Astronomy and Astrophysics, Peking University, Beijing 100871, P.R. China\\
	$^{3}$ Max-Planck institut f\"ur radioastronomie, Auf Dem H\"ugel, Bonn, 
	53121,Germany \\
	$^{4}$Bayesian Data Technologies (Wuhan) Co.,Ltd., Wuhan, Hubei 430079, China\\
	$^5$ Xinjiang Astronomical Observatory, National Astronomical Observatories, 
	Chinese Academy of Sciences, Urumqi, Xinjiang 830011, China\\
	$^{6}$Yunnan Astronomical Observatory, Chinese Academy of Sciences, Kunming 
	650011, China\\
	$^{7}$National Time Service Center (NTSC), Chinese Academy of Sciences,  Xi’an 
	City, Shaanxi Province, 710600\\
	$^{8}$CSIRO Astronomy and Space Science, Australia Telescope National 
	Facility, Box 76, Epping, NSW 1710, Australia\\
	$^9$ Radio Cosmology Laboratory, Department of Physics, Universiti Malaya, 
	Malaysia}
\date{Accepted 2021 March 16. Received 2021 March 16; in original form 2021 January 27}
\pubyear{2021}
\begin{document}

\label{firstpage}
	\pagerange{\pageref{firstpage}--\pageref{lastpage}}
\maketitle

\begin{abstract}
In this paper, we investigate the impact of correlated noise on fast radio burst 
(FRB) searching. We found that 1) the correlated noise significantly increases the 
false alarm probability; 2) the signal-to-noise ratios (S/N) of the false 
positives become higher; 3) the correlated noise also affects the pulse width 
distribution of false positives, and there will be more false positives with wider pulse width. We use 
55-hour observation for M82 galaxy carried out at Nanshan 26m radio telescope to 
demonstrate the application of the correlated noise modelling. The number of 
candidates and parameter distribution of the false positives can be reproduced with the modelling of correlated noise.  
We will also discuss a low S/N candidate detected in the observation, for which we demonstrate the method to evaluate the false alarm probability in the presence of correlated noise.
Possible origins of the candidate are discussed, 
where two possible pictures, an M82-harbored giant pulse and a cosmological 
FRB, are both compatible with the observation.
\end{abstract}
	
	
	
	
\section{Introduction}
Fast radio bursts(FRBs) are bright (50mJy-100Jy) millisecond-duration bursts 
observed in radio frequency as initially noticed by \citet{LBM07}.  The FRB 
signals, propagating through ionised medium, usually show frequency-dependent dispersion  features
following the cold-plasma dispersion relation, i.e. the pulses in high frequency 
band arrive earlier than those in low frequency.  The time delay between high and 
low frequency ($\nu_{\rm high}$ and $\nu_{\rm low}$) is $\Delta t=4.15\,{\rm 
ms}\, \times {\rm DM}
(\nu_{\rm low, GHz}^{-2}- \nu_{\rm high, GHz}^{-2})$, where DM is the electron column 
density along the line of sight in the unit of $\rm cm^{-3}\,pc$. The observed DM values of FRBs usually exceed those allowed by the Milky Way, which indicates the FRB sources are at extragalactic or cosmological distances. About 120 FRBs had been detected \citep{frbcat}\footnote{\textsc{frbcat}: \url{http://www.frbcat.org}}, and more than 20 of them reported to repeat \citep{FRB121102,CHIME2019a,CHIME2019b,Kumar2019,luo2020frb}. 
Recently, the 
origin of an FRB had been successfully traced to the magnetosphere of a magnetar 
\citep{LWM20, CHIMEsgr2020, BRB20}, yet the burst trigger mechanism is still 
unknown \citep{LZW20}.

FRBs have significant 
astrophysical applications, which cover a wide range of topics, e.g. testing the
Einstein's equivalence principle \citep{Wei2015PhRvL.115z1101W,
Tingay2016ApJ...820L..31T, Zhang2016arXiv160104558Z}, constraining
the rest mass of photons \citep{Wu2016ApJ...822L..15W,
Bonetti2016PhLB..757..548B, Bonetti2017PhLB..768..326B,
Shao2017PhRvD..95l3010S}, revealing hidden baryons in the
Universe \citep{McQuinn2014ApJ...780L..33M, MPM20}, studying the
dark-energy equation of states \citep{Zhou2014PhRvD..89j7303Z}, 
probing the
cosmological matter distribution \citep{Masui2015PhRvL.115l1301M}.

Finding a larger sample of FRBs is the key to understand the FRB burst trigger 
mechanism and to probe the astrophysics. The FRB searching is thus one of the 
most important observational activities in the field. Although FRBs are bright 
bursts, their short durations limit the signal-to-noise ratio (S/N). In order 
to detect FRBs, most of the FRB searching algorithms are to optimize the FRB 
detection in the presence of radiometer noise \citep{CM03,MLC19}. The matched 
filter being used in those investigations is the \emph{most powerful statistics} for detecting burst signals with 
the predetermined
waveforms \citep{VZ70}. It is \emph{asymptotically optimal} \citep{VV00}, i.e., it archives the 
best possible detection ability, when the signal-to-noise ratio is large. Indeed, several 
FRB searching softwares use box-car matched filter as the burst signal detectors, e.g.  
HEIMDALL \footnote{developed by Andew Jameson and Ben Barsdell,
\url{https://sourceforge.net/p/heimdall-astro}} and BEAR \citep{MLC19}. Two 
major key assumptions of the matched filter approach in signal detection are, 1) 
the waveform of the signal 
to be detected is known and 2) the statistical properties of the noise are known. The performance of the detector thus depends closely 
on the noise model.

One of the major obstacles in FRB searching is the interference in radio band.
The radio frequency interferences (RFIs) can mimic the FRB features in terms of the
intensity, pulse width and dispersion feature \citep{BBE11}. \citet{MLC19} had 
detected an RFI which resembles the FRB signal very well. Significant efforts had been 
applied to reduce the impact of RFIs on FRB searching. For example, methods such 
as RFI mitigation and machine learning in FRB candidate sifting, had both 
reduced the number of false positive detections \citep{ZGF18,AAB20,ZWH20}. For strong RFIs, one
usually mitigates the effects by subtracting them from data. Weak RFIs, on the 
other hand, are hard to be subtracted. Hence, the correlated noise (or coloured noise) will be left in the data.

The correlated noise is well known in radio astronomy, and it may 
emerge from many different processes, e.g., flicker noises \citep{press78}, 
propagation effects \citep{GJ11}, signal-chain gain instability \citep{JLI04}, and 
low level RFIs \citep{Dewey94}.  Investigating the origin of the correlated 
noise in FRB searching data is beyond the scope of this paper. Here we focus on 
the impact of the correlated noise on FRB searching. 

We will 
show in 
\SEC{sec:stat} that the correlated noise increases the false alarm probability for FRB detection dramatically, and the distribution function of the burst parameters will also be
affected. We will investigate the S/N and pulse width 
distribution of the false positives.  We use real observational data to illustrate 
the application of the correlated noise modelling in \SEC{sec:application}, where we 
also discuss the evaluation of false alarm probability for single event, where
a potential FRB candidate found during the observation of the M82 galaxy is used as an example.  
Discussions and conclusions are made in \SEC{sec:con}.

\section{Detection statistics}
\label{sec:stat}

One can show \citep{MLC19} that the optimal statistic $S$, in the sense of the \emph{asymptotically most powerful test} \citep{VV00}, for detecting the pulse 
signal of square waveform from the background of uncorrelated Gaussian noise is
\begin{equation}
S=\frac{1}{N_{\rm box} \sigma^2}\left(\sum_{|t-t_0|<w} s_i\right)^2\,,
\label{eq:det}
\end{equation}
    where $N_{\rm box}$ is the number of data points within the time span of the burst, 
    specified by pulse width $w$, pulse epoch $t_0$, and time $t$ with 
$|t-t_0|<w$,  $\sigma$ is the root-mean-square level of the Gaussian noise. With such a statistic, one will claim that a burst is detected, if the statistic $S$ is greater than a pre-set threshold $S_0$. 
The null distribution of $S$, i.e. the distribution of $S$ when the data contains only noise 
but no bursts, follows the one degree of freedom $\chi^2$ 
distribution \citep{CM03,MLC19}. The false alarm probability, i.e.  
the probability of reporting a `detection' due to statistical fluctuation 
when no real burst happens, becomes
\begin{equation}
	P_{\rm f}(S\ge S_0)={\rm erfc}(\sqrt{S_0/2})\,.
	\label{eq:fap}
\end{equation}
Here, function ${\rm erfc}$ is the complementary error function.
This statistic is constructed under a rather ideal condition, where only 
the uncorrelated Gaussian noise is assumed. When we include coloured noise, 
i.e. noise with temporal correlation, the statistical distribution of $S$ 
differs from the above result. As shown in \APP{sec:redS},
the detection statistic $S$ follows the `scaled' one degree of freedom $\chi^2$ 
distribution. Both the mean and standard deviation of statistic $S$ increase.  
The revised false alarm probability affected by the correlated noise becomes
	\begin{equation}
	P'_{\rm f}(S\ge S_0|\eta)={\rm 
		erfc}(\sqrt{S_0/2\eta})\,.
	\label{eq:fapt}
	\end{equation}
	We find that the scale factor $\eta$ can be approximated by
	\begin{equation}
	\eta=1+{N_{\rm box}\kappa}\frac{\sigma_{\rm r}^2}{\sigma^2+\sigma_{\rm r}^2}\,,
	\label{eq:scale}
	\end{equation}
	with $\kappa$ being a numeric factor depending on the noise spectral shape and 
	$\sigma_{r}$ being the root-mean-square (RMS) level of the coloured noise. For 
	power-law noise, as explained in \APP{sec:redS}, $\kappa\simeq 1/100$.
	
\begin{figure}
		\includegraphics[width=3.1in]{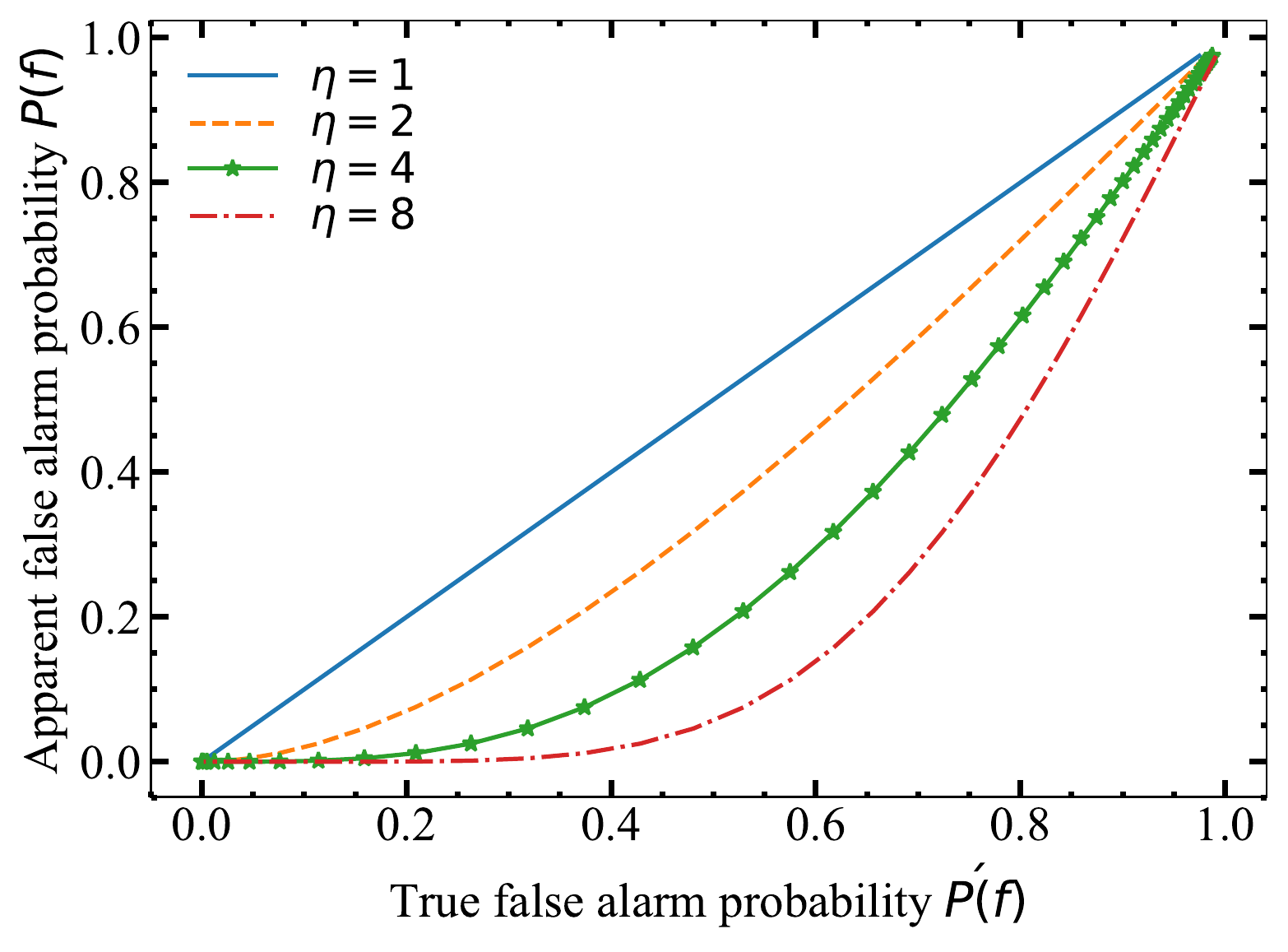}
		\caption{The relation between the true false alarm probability $P'_{\rm f}$ (\EQ{eq:fapt}) and the apparent false alarm probability $P_{\rm f}$ (\EQ{eq:fap}). The scale factor $\eta$ is labelled in the figure.  
			\label{fig:pfcmp}}
	\end{figure}

The relation between white-noise-only false alarm probability $P_{\rm f}$ and 
the one including the coloured noise component is plotted in \FIG{fig:pfcmp}. One can 
see that the false alarm probability computed using the \EQ{eq:fap} will be 
underestimated, if the signal contains coloured noise. The larger $\eta$ is, the 
more one will underestimate the false alarm probability. That is, a high value 
of $S$ may come from correlated noise contribution rather than from real burst 
signals.

As shown in \EQ{eq:scale}, two factors play the roles, 1) the amplitude of the 
coloured noise ($\sigma_{\rm r}$) and 2) the number of data points ($N_{\rm 
box}$) within the given pulse. One can see that a higher RMS level of coloured noise 
or a wider pulse introduces a larger bias on the false alarm probability 
computation.  $N_{\rm box}$ in \EQ{eq:scale} reflects the fact that a wider 
pulse (corresponding to a lower frequency in frequency-domain spectrum) contains 
more coloured noise contributions.  For a typical case, the data sampling time scale 
is about 100 $\mu$s and the FRB signal lasts for a few milliseconds, $N_{\rm 
box}\sim 10-10^{2}$. Considering $\kappa\simeq 0.01$, the red noise component 
plays a significant role, when its RMS level is compatible with the white noise.
	
The correlated noise also changes the \emph{finite-sample false alarm 
probability}, which is defined as the false alarm probability for all candidates 
with all possible parameter combinations.
With coloured noise, the false alarm probability of a single event depends on $\eta$. The 
finite-sample false alarm probability ($P_{\rm }(S\ge S_0)$)  is found by 
integrating over all possible value of $\eta$, i.e.
	\begin{equation}
	P_{\rm f}(S\ge s_0)=\int P'_{\rm f}(s\ge s_0|\eta) f_{\rm \eta}(\eta) d\eta\,
	\label{eq:allpf}
	\end{equation}
where $f_{\rm \eta}$ is the probability density of $\eta$.  We have $f_{\eta} \propto T/N_{\rm box}$, if the data length is fixed to be $T$.

	\section{A representative example}
	\label{sec:application}
\subsection{Observation and data reduction}
We now show the application of the above theory. The data for the 
demonstration was taken with Nanshan 26-meter (NS26m) radio telescope, which was 
observing the M82 galaxy at that time. The NS26m is operated by Xinjiang Astronomical 
Observatory (XAO) of Chinese Academy of Science.  Its location is N$43^\circ 
28.27'$, E$87^\circ 10.67^\prime$ with the altitude of 2080m 
\citep{wang2001pulsar}. We perform observations centred at 1560 MHz with 
bandwidth of 320 MHz, i.e. from 1.4 GHz to 1.72 GHz. NS26m has a cryogenic 
frontend and the total system temperature is 25 K. The radio frequency signal is 
down converted to intermediate frequency of 100-420 MHz with a local oscillator 
at 1300 MHz. 
	
We also used data collected from the Kunming 40-metre (KM40m) and the Haoping 
40-meter (HRT)
radio telescopes. KM40m is operated by Yunnan Astronomical Observatory Chinese 
Academy of Sciences. It locates at N$25^\circ 02'$, E$102^\circ 47^\prime$
with the altitude of 1985 m \citep{HWY10}. We use C-band (~ 6.7 GHz) data of 
KM40.  The HRT at Haoping  (N$34^\circ, 10.5'$ E$109^\circ 56.97^\prime$) 
\citep{LGY20}, is operated by National Time Service Center, Chinese Academy of 
Sciences. The electronic specification and observation configuration of the 
telescopes are given in \TAB{tab:parameter}.

	We use the \textsc{Reconfigurable Open Architecture Computing Hardware 
		2}\footnote{\textsc{roach2}: \url{https://casper.ssl.berkeley.edu/wiki/ROACH2}}  
	based system to digitize the signal at NS26m, where we form 1024 channels using 
	polyphase filter and integrate the intensity to form filterbank data with 65 
	$\rm{\mu s}$ sampling time. The filterbank data packets are then transferred to a data 
	recording computer using 10 Gigabit Ethernet. 
	
	\begin{table}
		\centering
		\caption{Setups at NS26m, KM40m and HRT for M82 observation \label{tab:parameter}}
		\begin{tabular}{lcccccr}
			\hline \hline
			Telescope & BW& $f_{\textrm{central}}$ $^1$ & $f_\textrm{ch}$ $^2$  & Gain & 
			$T_\textrm{sys}$ $^3$ & $\Delta t$ $^4$ \\
			& MHz & MHz & MHz & $\rm{K\,Jy}^{-1}$ & K & $\rm \mu s$\\
			\hline
			NS26m &320 & 1560 & 0.97 & 0.1 & 25 & 65\\
			KM40m &440 & 6690 & 0.2 & 0.2 & 96 & 65\\
			HRT &800 & 1400 & 0.097 & 0.2 & 100 & 163\\
			\hline \hline
		\end{tabular}
		\raggedright
		
		$^1$ Central frequency of observation\\
		$^2$ Channel width\\
		$^3$ System temperature \\
		$^4$ Time resolution
	\end{table}
	
We observed M82 with NS26m for five times from 2016 to 2017. The total observing 
time is 55 hours. The FRB searching is performed in realtime 
using the software package \textsc{Burst Emission Automatic 
Roger}(\textsc{BEAR}, \citealt{MLC19}).  In \textsc{BEAR}, radio frequency 
interference (RFI) mitigation, de-dispersion, box-car matched filter pulse 
detection, and candidate clustering are performed.  There are two RFI mitigation 
steps in our pipeline: we first remove data of the channels around 1.55 
GHz, because of the RFI induced by satellite communication, and  then we use the 
zero-DM matched RFI filter \citep{MLC19} to remove wideband interference 
without dispersion features. We further neglect all candidates with DM bellow 
200 $\rm cm^{-3}\, pc$ in our analysis to remove the zero-DM RFI contamination.

The data was downsampled before de-dispersion to reduce the computational cost.  
The parameter of downsampling is tuned together with the DM steps in 
de-dispersion.  We choose the largest possible DM step such that the pulse 
smearing induced by DM-mismatching is less than the inter-channel DM smearing.  
The downsampling time resolution is chosen such that it is also smaller than the 
inter-channel DM smearing. In this way, we minimize the computational cost 
without downgrading the minimal width of detectable pulses, and the smearing 
effects are mainly affected by the inter-channel DM smearing.  
	
\begin{figure}
		\includegraphics[width=3.5in]{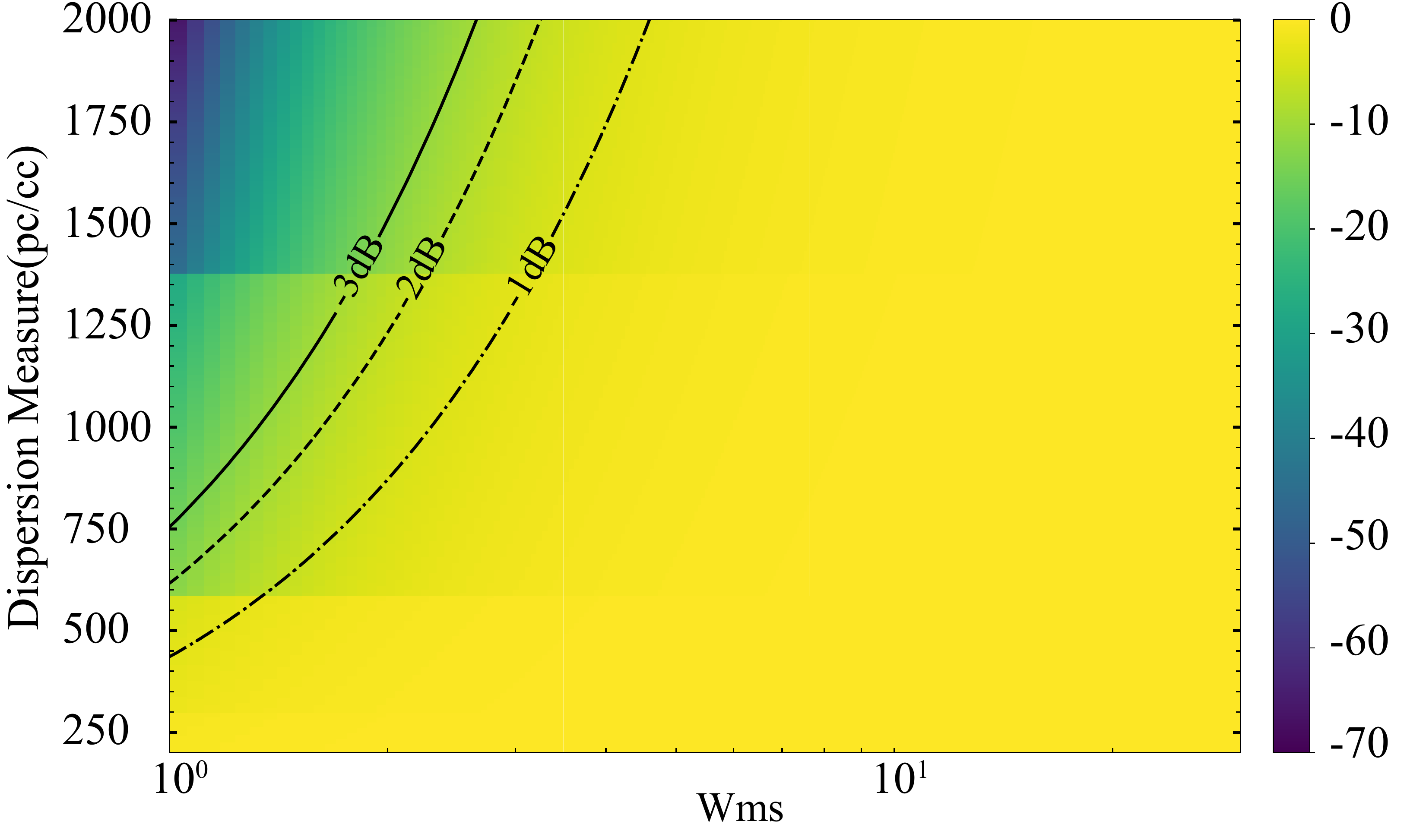}
		\caption{The S/N loss curve caused by sub-band incoherent de-dispersion 
		process. Here our computation uses NS26m parameters. The S/N loss includes 
		effects of inter-channel DM smearing, searching DM step size, and 
		downsampling. X-axis is the pulse width in ms, and Y-axis is the DM of the 
		source. The S/N loss is given in dB, i.e. $10\log_{\rm 10}\left(\rm{S/N} 
		\right)$.The contour lines represent the total S/N loss for 1, 2 and 3 dB.}
		\label{fig:snrloss}
	\end{figure}

Subband de-dispersion was used in \textsc{BEAR} to speed up the computation, 
where we divided the filterbank data into subbands and de-dispersed the data 
into a coarse grid of trial DMs. The de-dispersed subband data are then further 
de-dispersed to the required fine DM grids \citep{MLC19}. We chose DM range from 
200 to 2000 $\rm cm^{-3}\,pc$. Our plan of de-dispersion is listed in 
\TAB{tab:ddplan}.  The calculation of the S/N loss is shown in 
\FIG{fig:snrloss}, where one can see that our setup is sensitive to FRBs with 
widths larger than 3 ms for  ${\rm DM} < 2000 {\rm cm^{-3}\, pc}$.	
\begin{table*}
		\centering
		\caption{The de-dispersion plan for data of NS26m. \label{tab:ddplan} }
		\begin{threeparttable}
			\begin{tabular}{cccccccc}
				\hline\hline
				DM range & $\Delta \textrm{DM}$ $^1$ & Downsampling & $\Delta 
				\textrm{sub}_\textrm{DM}$ & Subband  & $\tau_{\rm ch}$ $^3$& $\tau_{\rm \Delta 
					DM}$ $^4$\\
				($\rm cm^{-3}\,pc$) &($\rm cm^{-3} pc$)  & ratio &($\rm cm^{-3}\,pc$) & number  & (ms)& (ms) \\
				\hline
				200.0 - 300.5 & 0.5& 4:1 &  8& 12 & 0.3   &0.4\\
				300.5 - 591.5 & 1.00 &8:1 &17 & 17 & 0.6  &0.7\\
				591.5 - 1389.5 & 3.00 &16:1 &18 &45 & 1.2  &2.0\\
				1389.5 - 2004.5 & 5.00 & 32:1 &18  &34 &2.0 &3.0\\
				\hline \hline
			\end{tabular}
			\raggedright
			$^1$ DM step size \\
			$^2$ Subband DM step size\\
			$^3$ Maximal inter-channel DM smearing time scale\\
			$^4$ DM-mismatch smearing time scale\\
		\end{threeparttable} 
	\end{table*}
	
\begin{figure*}
		\includegraphics[width=\textwidth]{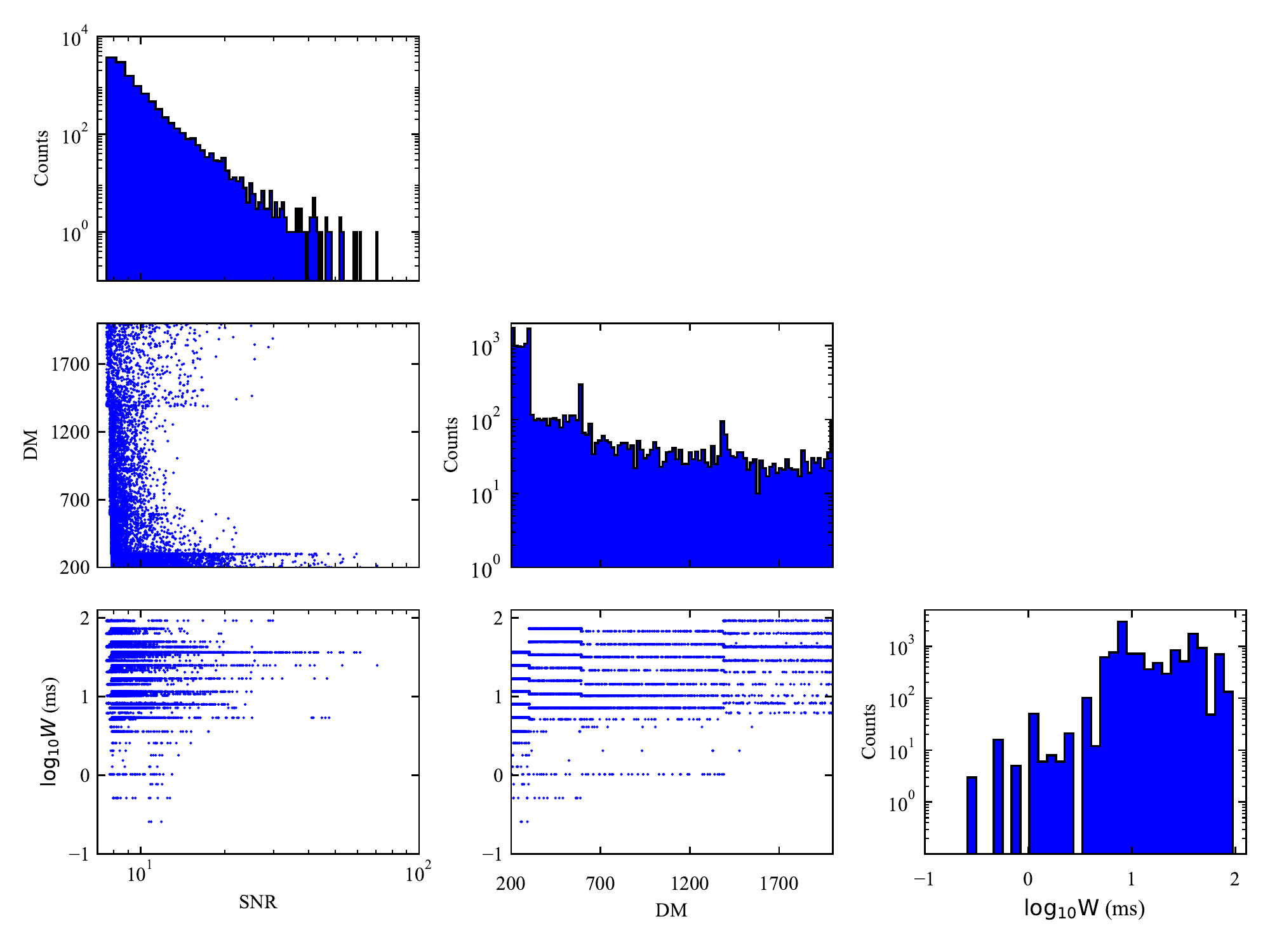}
		\caption{Candidate distribution of NS26m for DM, pulse width 
		$W_\textrm{ms}$ and S/N. Here we only include the candidates with ${\rm 
		S/N}\ge 7$. The diagonal histograms are one-dimensional distributions for 
		each parameter, i.e. S/N, DM, and logarithmic pulse width. The off-diagonal 
		scatter plots are for the two-dimensional distributions of parameter pairs.
		A few peaks in DM distribution, i.e. around ${\rm DM}\simeq300, 600$, and $1400$, are due to our de-dispersion plan as shown in \TAB{tab:ddplan}. On the boundaries of the de-dispersion plan, bursts were searched `twice', as the de-dispersion plan segmentation overlaps. Furthermore, if the true DM of a burst is slightly outside the de-dispersion range, the searched DM value is forced to be on the boundary. In this way, high S/N candidates further pile up on the DM boundaries. 
			\label{fig:pardis}}
	\end{figure*}

After the candidate plots are generated, the plots are visually inspected to see 
if there is a wideband burst signal with a dispersive signature.  There are about 
$1.4\times 10^4$ candidates generated with ${\rm S/N}\ge 7$. The 
multi-dimensional distributions for the candidate parameters S/N, width, and DM
are shown in \FIG{fig:pardis}. Four notable features in \FIG{fig:pardis} are 
topic-related for the discussion later in the paper, 1). the candidate number is 
much larger than expected; 2). the candidate S/N distribution shows long-tailed feature; 3). candidate counts correlate with the pulse width; 4). S/N correlates with the pulse width.  	

\begin{figure}
		\includegraphics[width=3.5in]{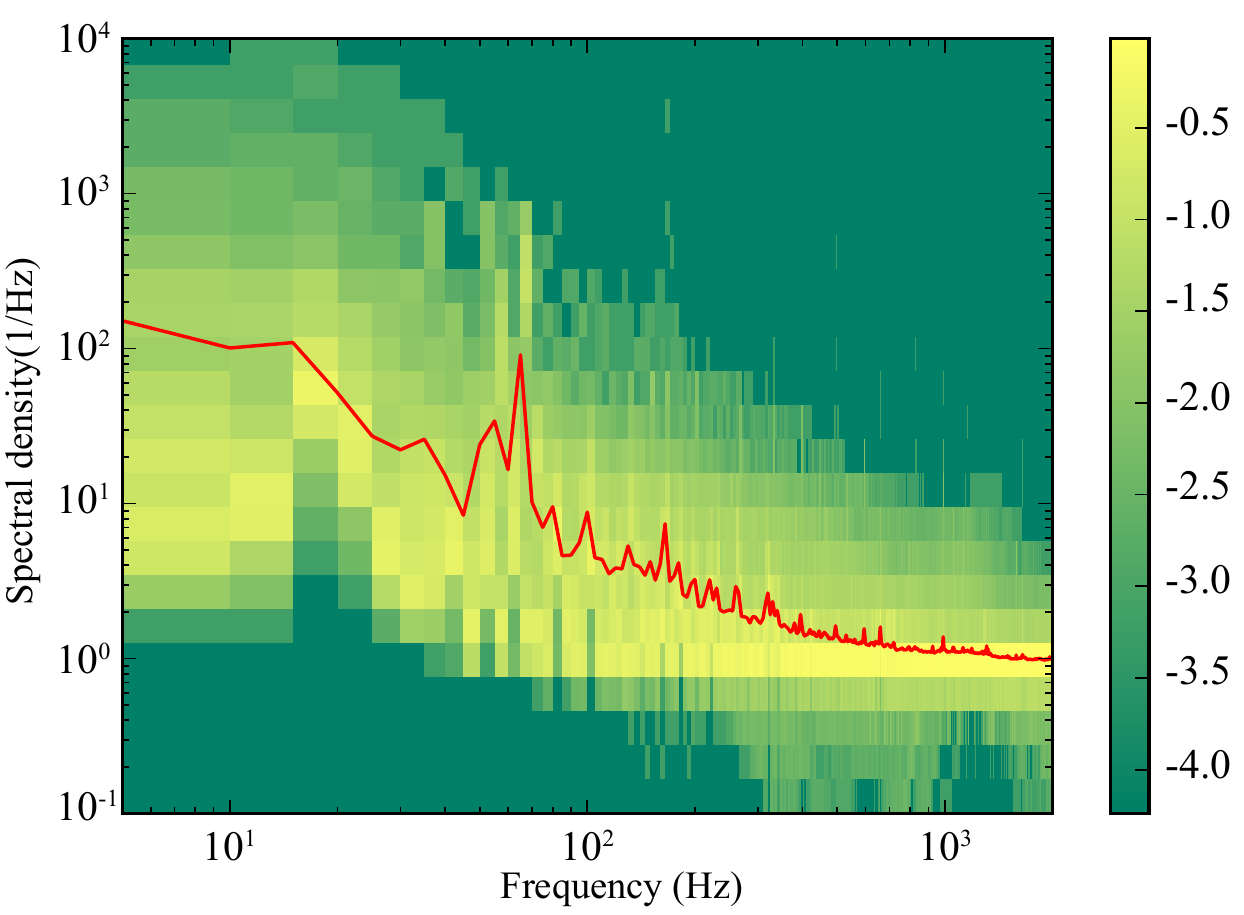}
		\caption{The noise spectrum of 0-DM time series. The x-axis is the frequency in Hz, and the y-axis is the spectral density estimated using Hamming windowed Fourier transform. The red solid curve is the average of all spectra. The background colour scale is the 10-based logarithmic probability of occurrence for the noise power as the given frequency and spectral density. The colour bar on the right side indicates the probability. \label{fig:spectrumall}}
	\end{figure}

Those four features are unexpected in white Gaussian noise modelling \citep{CM03,MLC19}. To understand the four features, we need to characterise the correlated noise. We 
measured the correlated noise spectrum by performing windowed spectral analysis 
on zero-DM one-dimensional time series. The data was divided into a number of
one-second segments, and the Hamming windowed Fourier transform \citep{Harris78} was 
applied to estimate the spectral density. The spectral density of all 
the data is collected in \FIG{fig:spectrumall}. As one can see that the 
average spectrum is \emph{dominated} by low-frequency correlated noise, and the spectral density drops as the frequency increases, only at frequency above a few kHz, the spectral density curve gradually turns flat. Obviously, the false alarm probability will be underestimated, if 
we use the white-noise-only model. 

The intensity and shape of red noise spectrum fluctuate time-to-time, as 
indicated by the background colour shade in \FIG{fig:spectrumall}. 
The data can be `white' or `red' for a short duration. Since, by \emph{average}, the red noise dominates the noise spectrum, we have $\eta \simeq 1+N_{\rm 
box}\kappa\sim 1+0.01 N_{\rm box}$.
	
\begin{figure}
\includegraphics[width=3.5in]{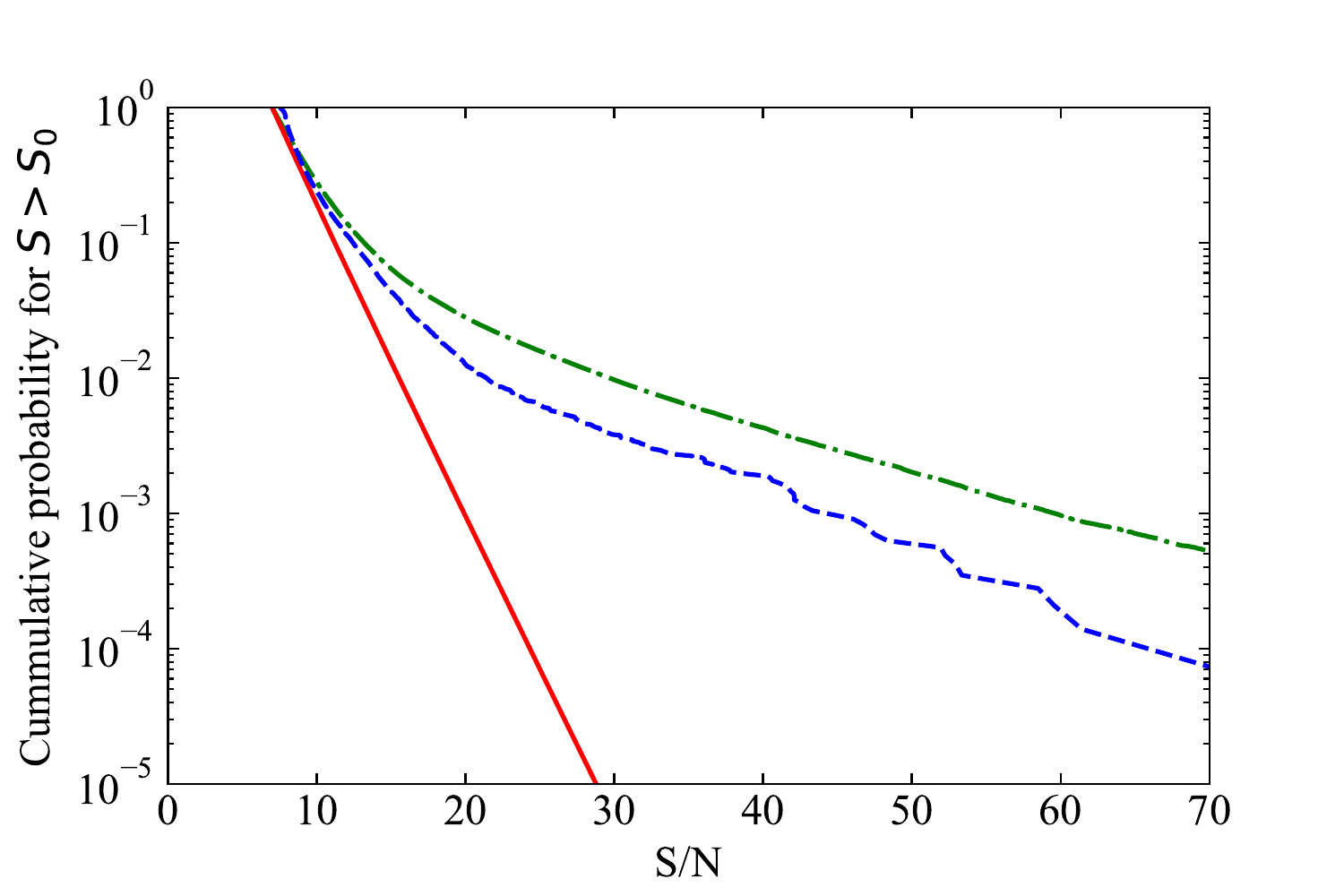}
\caption{The relation between S/N and cumulative probability for candidates with 
$S\ge S_0$. The x-axis is the S/N, i.e. $\sqrt{S}$, and the y-axis is the 
cumulative probability for $S\ge S_0$. The measured cumulative distribution is 
plotted in blue dashed curve. The red solid curve is computed from the $\chi^2$ 
distribution, i.e. \EQ{eq:fap}. The green dash-dotted curve are computed from 
\EQ{eq:allpf}, which includes the correlated noise contribution.  
\label{fig:snrdis}}
\end{figure}

We can check the distribution of all-candidate S/N as shown in \FIG{fig:snrdis}.  
The distribution of detected candidates deviates from the $\chi^2$ distribution.  
Because the expected $S$ is higher for pulses with larger width. The wider 
pulses will induce a higher S/N distribution tail. The measured distribution 
indeed has a tail extending to higher S/N. The false alarm probability will be 
underestimated, if one assumes $\chi^2$ distribution, i.e.\EQ{eq:fap}. The 
correct version of false alarm probability can be computed by including the 
correlated noise modelling, i.e. by using \EQ{eq:allpf}.  As one can see in 
\FIG{fig:snrdis}, \EQ{eq:allpf} produces S/N distribution similar to what 
see in the data. 

We can further see that the false alarm probability calculated using \EQ{eq:fap} 
does not agree with the number of candidates. If only white noise is assumed, 
${\rm S/N}\ge 7$ is equivalent to $S\ge 49$, and one gets the false alarm 
probability of $P_{\rm f}\simeq2.6\times10^{-12}$ according to \EQ{eq:fap}. It 
is contradicting that we detected approximately $1.4\times 10^{4}$ 
candidates for 55 hours observation, where the expectation 
detection number 
is about $N\simeq 2\times 10^{-4} (P_{\rm f}/10^{-12}) (T/55 {\rm h}) (w/{\rm 
ms})^{-1}$. If we include the coloured noise modelling, the expected number of 
candidates $N_{\rm c}$ is
    \begin{equation}
	N_{\rm c}= \sum_{w_{i}} P_{\rm f }(S\ge 7|\eta(w_i))\frac{T}{w_{i}}\,,
	\end{equation}
	where the summation is performed over all pulse width $w_{i}$ used in the pulse searching. The above equation gives $N_{\rm c}\simeq1.5\times 10^4$, which matches what we see in real data.

\begin{figure}
		\includegraphics[width=\columnwidth]{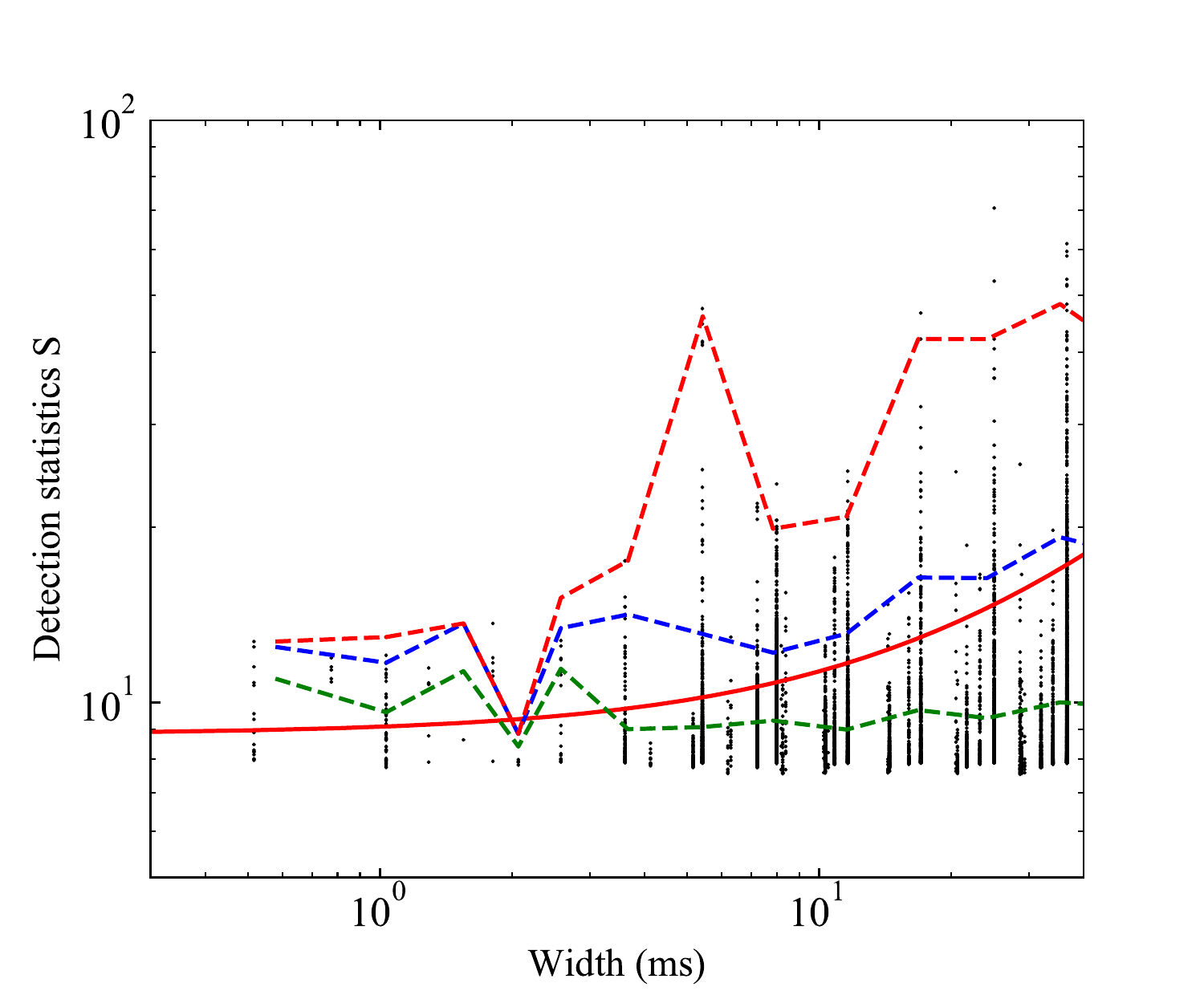}
		\caption{The relation between pulse width (x-axis) and detection statistic (y-axis). The black dots are for all candidates found in observation. The solid line in green shows the average value of the detection 
			statistic as a function of pulse width, while the solid blue and red curve are the 95\% and 99.7\% envelope. The corresponding curves in dash lines are the same curve computed from red noise modelling, i.e. from \EQ{eq:allpf}. 
			\label{fig:snrw} }
	\end{figure}

We also note that the pulse width is correlated with the detection statistic $S$, 
which reflects the coloured noise effect. As shown in \FIG{fig:snrw}, there 
seems to be an increase of S/N with larger 
pulse width. For pure white 
noise modelling, the correlation is not expected.  After including the coloured 
noise, i.e. \EQ{eq:allpf}, we can understand such a correlation.
	
\subsection{An interesting candidate}

For 55 hour data, all candidates with S/N$\ge5$ were visually inspected. Particularly, volunteers had
helped to perform the visual inspection of all candidates more than
once. We organised the visual inspection campaign by collaborating
with \emph{Bayesian Data Technologies (Wuhan) Co., Ltd} (BDT). BDT
helped to remove the axes of candidate figures, and embedded the
candidate plots into their mobile game \emph{Lighthouse Project}, aiming at popularizing
science\footnote{\url{http://www.bayesiandt.com/}}.  The players of the game are 
encouraged to acquire
more credits by identifying the contents of figures. In the game,
the players need to first pass a training session.  In the session,
examples of known FRB signals were shown to the players together
with explanation of the contents. Common concepts in astronomy are
introduced, e.g. dispersion, frequency, and intensity.
In this
way, the players can understand what were shown to them.  After the 
training, candidate plots were shown, and the players started to
help identify the FRB signals.  Each candidate plot was passed
to several players, and BDT helped to identify the common votes
from players. We 
also tested whether there are candidates being
missed by the volunteers. This is done by showing known FRB signals
to the players, and the `performance' of each player can be evaluated
by computing the probability of missing true FRBs.

\begin{figure} \includegraphics[width=\columnwidth]{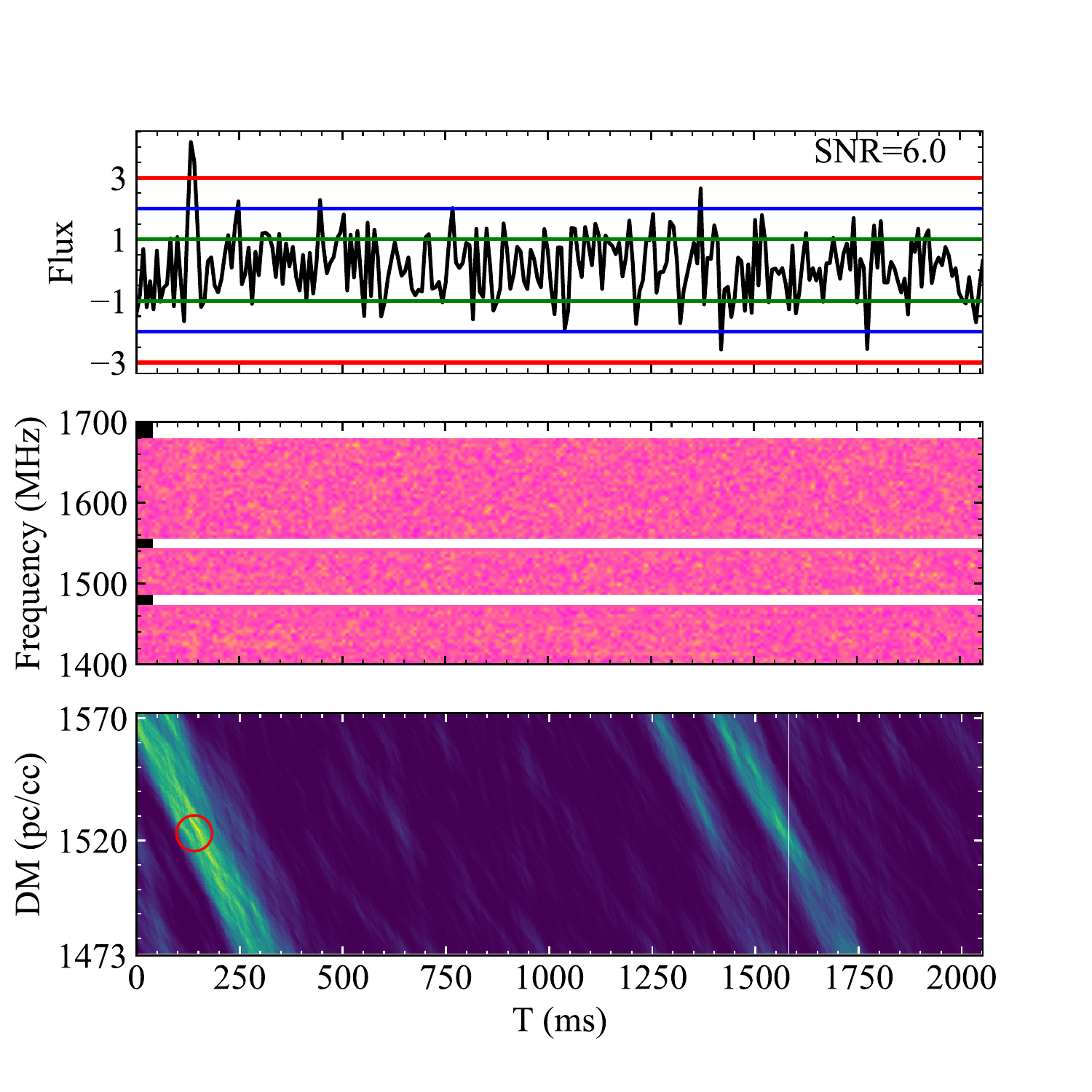}
	\caption{Dynamical spectra of a weak burst (${\rm S/N}\simeq6$, after RFI zapped) detected with 
	NS26m
telescope (320MHz).  Top: normalised  pulse profile. The pulse width
is approximately 9 ms
Middle: dynamical spectrum of the pulse. 
The pulse has been de-dispersed to DM 1522.8
$\textrm{pc } \textrm{cm}^3$ to get maximal ${\rm S/N}=6$. 
We reduce the number of frequency channel to 82 (channel
width of 3.9 MHz) and time resolution to 1.15 ms, the total effective bandwidth is 280 MHz. After de-dispersing with cold-plasma dispersion relation, 
i.e. $f^{-2}$-law, the signal in each channel are aligned in time.
Bottom: S/N as a function of time and DM trials. The red circle indicates the pulse position in DM-time parameter space. }

\label{fig:pulse} 
\end{figure}

With the help of volunteers, we also looked at those weaker bursts
with $5\le{\rm S/N}\le7$. We found one interesting candidate within
the entire 55 hour NS26m observation. The pulse profile, de-dispersed dynamical
spectra, and S/N-DM-t diagnosis plot of the candidate is shown in \FIG{fig:pulse}. More details can be found in \APP{sec:bearplot}. The candidate
exhibits wideband emission, `-2'-index DM signature, but with low
S/N. Those properties make it hard investigate the origin of the candidate. It can be
either a RFI or a real FRB. We estimated the flux and fluence of the burst to
be 0.6 Jy and 7 ${\rm Jy\, ms}$, using the gain ($0.1{\rm K/Jy}$), system temperature (25 K) of NS26m, and effective bandwidth of 280 MHz after RFI mitigation.

\begin{figure}\includegraphics[width=\columnwidth]{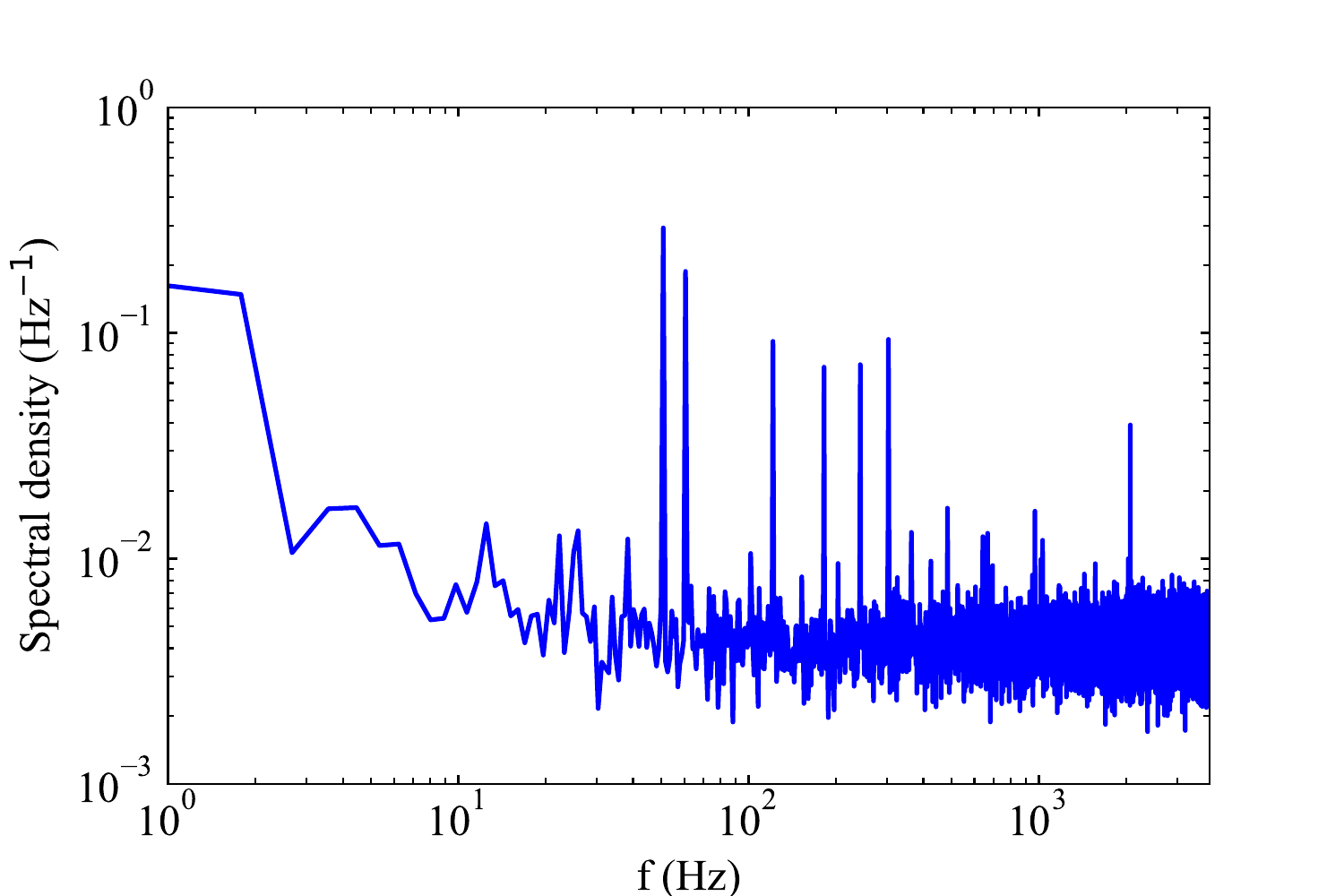}
		\caption{The noise spectrum of the data shown in \FIG{fig:pulse}. The x-axis 
		is the frequency, and the y-axis is the power spectrum density.  Besides the 
		white noise components (horizontal part with frequency higher than 100 Hz), 
		there is a low frequency red noise components together with several periodic 
		components, which may come from RFIs. By subtracting the white noise 
		component, we measured that the ratio between red noise power and total 
		noise power is $\sigma_{\rm r}^2/( \sigma^2+\sigma_{\rm r}^2)\simeq6\%$, 
		with sampling time of 130 $\mu s$. 	\label{fig:noispec} }
	\end{figure}

The pulse may come from statistical fluctuations. The S/N of the pulse 
is 6.0, which corresponds to $S=36$ and $P_{\rm f}\simeq 2\times 10^{-9}$ for pure white noise case. Given 
the pulse width of 9 ms and total 55 hour observation time, there will be $2.2 
\times 10^7$ independent 9-ms segmentations. Thus, if only the white noise is 
considered, one will expect the chance to find such a `burst' due to the 
\emph{accidental} statistical fluctuation is about $1\%$. 
As we have shown in the previous section, red noise would
increase the apparent $S$ and significantly increase the false alarm probability.
Could the candidate be caused by the red noise then? We extracted eight-second data around the 
burst, and measured the noise power spectrum of zero-DM time series. The 
spectral density is plotted in \FIG{fig:noispec}. We estimate the  white noise contribution 
by fitting a horizontal line
for frequency 
above 100 Hz. We then measured the red noise contribution, by subtracting the 
white noise component from the total spectrum. For the 8-second data around the 
pulse, the red noise component contribution is only about $6\%$ of total power, i.e.  $\sigma_{\rm 
r}^2/(\sigma^2+\sigma_{\rm r}^2)\simeq 6\%$. The data around this pulse 
seems to be less affected by red noise. According to \EQ{eq:scale}, red noise is 
incapable of affecting the $S$ estimation significantly. We should be able to trust 
the 6-$\sigma$ ${\rm S/N}$ and 1\% false alarm probability for 55-hour 
observation.  From probability point of view, it is preferred that the burst is 
real and not from statistical fluctuation.

Follow-up observations for the source were carried out using KM40m and HRT. We 
observed 12 hours with KM40m on 8th Nov. 2019 using 6.7 GHz center frequency.  
and 12 hours with HRT on 30th Dec. 2019 at 1.4 GHz center frequency.
Unfortunately, neither HRT nor KM40m detected any convincing pulse in the DM range from 1520 to 1526 $\rm cm^{-3}\,pc$ with S/N$\ge 6$, and the flux limits 
are $\le 1.4{\rm Jy\, (1.4 GHz)}$ and $0.7{\rm Jy\, (6.7 GHz)}$ respectively, 
assuming the pulse width of 9 ms as measured by NS26m.

\section{Discussion and conclusion}
\label{sec:con}
In this paper, we investigate the effects of correlated noise in the false 
alarm probability computation for FRB detection problem. In the presence of
correlated noise, particular red noise, the false alarm probability becomes 
larger than the case where only white noise is included. The correlated noise 
also introduces a dependence of false alarm probability on the pulse width.  
Particularly, for the red noise case, one will detect more candidates with 
larger pulse width. We have also used observation carried out at NS26m 
radio telescope to show the application of the false alarm probability 
computation. We can reproduce the expected number of candidates and their 
statistical properties. We also show one interesting candidate found in M82 observation.

The burst appears in the direction of M82, however we can not
directly associate the burst with M82. The beam size (in solid angle) of
NS26m is roughly 16 times of the angular size of M82 (approximately
11.6' $\times$ 3.7' \citep{jarrett20032mass}). In fact, NASA/IPAC
Extraglaxtic Databases \footnote{\url{https://ned.ipac.caltech.edu}} show that
there are 1458 known galaxies in the NS26m field of view.
If we assume the source resides in M82, given the
luminosity distance of 3.6 Mpc for M82, the isotropic
peak luminosity of the burst will be $L_{\rm iso}\sim
3\times 10^{36} \textrm{erg/s}$, which will be four
orders of magnitude lower than the average  luminosity of
known FRBs. The luminosity is compatible with that of
bright Crab giant pulses reported by \citet{BC19},
although the pulse width is at least three orders of
magnitude wider than the case of the Crab giant pulse.
The luminosity is also similar to the case of
SGR~J1935+2145 ($\sim 7\times 10^{36} \textrm{erg/s}$
\citealt{CHIMEsgr2020,BRB20}), although the 
pulse width is about 20 times larger than that of SGR~J1935+2145 (i.e. 0.5 ms \citealt{CHIMEsgr2020}).
If the burst is real, and it is located in M82, it is compatible to the picture that a bright radio burst comes from a magnetar in the M82.

The measured DM (1523 $\rm cm^{-3}\,pc$) of the candidate is compatible with both of the two scenarios, 1) a giant pulse from M82, and
2) an FRB at a cosmological distance. The DM contribution from the intergalactic medium between Milky Way and M82 ($\sim 1\,\rm cm^{-3}\,pc$, \citealt{frb-halo})  is negligible comparing to other contributions, i.e.  Galactic foreground of 33 $\rm cm^{-3}\,pc$ \citep{yao2017new}, halo of local group $\sim 100\,\rm cm^{-3}\,pc$ 
\citep{frb-halo}. Thus, the major DM contribution (1400 $\rm cm^{-3}\,pc$ ) results from M82, if M82 association is assumed. As shown by
\citet{westmoquette2007hubble} and \citet{heckman1990nature}, the electron density in M82
increase from 100 $\rm cm^{-3}$ at 1-3 kpc outskirt to
1000 $\rm cm^{-3}$ around the galaxy nucleus. 
Therefore,
the measured DM is compatible with M82 properties in the
picture of \emph{an M82 hosted source}.
The other picture, i.e. \emph{a cosmological FRB} is also possible. If the source 
is at a further cosmological distance, the source can be much more luminous.  
Using the model of \citet{luo2018normalized}, the redshift will be 
$z=1.3^{+0.03}_{-0.5}$ and the inferred isotropic peak luminosity is 
$1.9^{+0.1}_{-1.3} \times 10^{43}\,{\rm erg/s}$, which is compatible to the 
properties of the known FRB population \citep{luo2018normalized}.  At this 
stage, we 
are lack of further detection, we can not conclude yet on the nature of the 
source, as both the M82 and the cosmological interpretation are possible.

In both of the two pictures, there is an event rate issue.  If the 
burst is harbored by M82,  the \emph{inferred} event rate is $1/55 \simeq 0.02\, 
{\rm ~hour^{-1}}$.  Considering the rareness of radio burst from the Galactic 
magnetar SGR~J1935+2145 and the fact that M82 are smaller in stellar 
number comparing to the Milky Way, the inferred event rate value seems to be too 
high.  If the burst is cosmological, we would expect to detect $10^{-2}\sim 10^{-3}$ FRB per 
day with NS26m \citep{luo2020frb}, which is still much lower than our inferred 
event rate.  However, the star formation rate in M82 is a few times higher 
than that in the Milky Way\citep{KE12}, which may indicates that the event rate of radio 
burst phenomenon may correlate with the star formation rate.  If so, we would 
expect more radio bursts can be found by monitoring those star burst galaxies.

\section*{Acknowledgements}
The work is supported by CAS XDB23010200, Max-Planck Partner Group, National SKA program of China 2020SKA0120100, NSFC 
11690024, CAS Cultivation Project for FAST Scientific. We also thank active players of the mobile game \emph{Lighthouse Project} to help perform the candidate inspection, including Y. Zou, H.~Y. Lin, J.~Q. Li, J.~Y. Li, Y.~X. Jiang, etc.
\section*{Data Availability Statement}
The data underlying this article are available in a repository and can be accessed via \url{https://github.com/zhanyige/M82_FRB_couloured_noise/blob/data/original_data.zip}
	
	
	
\bibliographystyle{mnras}
\bibliography{ms} 

	
	
	
	
\appendix
	
\section{Statistical properties of S with coloured noise}
\label{sec:redS}
	
We denote the coloured noise component as $r_i$, and white noise component as 
$n_i$ with index $i$ indicating the temporal sampling. The de-dispersed one 
dimensional time series is $s_i=r_i+n_i$.
	The summation of $s_i$, i.e. $u\equiv\sum_{|t-t_0|<w} s_i$, follows Gaussian distribution, because a linear superposition of Gaussian variable is still Gaussian. In this way, the detection statistic, being a square of Gaussian ($S\propto u^2$), must follow a one dimensional scaled $\chi^2$ distribution.
	
We can compute the mean and standard deviation to fully determine the 
distribution. By expanding the correlation, one can show 
that
	\begin{eqnarray}
	\langle S\rangle&=& 1+\frac{1}{N_{\rm box}}\sum_{i,j} \gamma_{ij}  \,,\\
	\langle S^2\rangle-\langle S\rangle^2&=&2\left( 1+\frac{1}{N_{\rm box}}\sum_{i,j} \gamma_{ij}\right)^2 \,.
	\end{eqnarray}

Here $\langle \cdot\rangle$ indicates the ensemble average. $\gamma_{i,j}$ is 
the two-point correlation of coloured noise normalised by the total noise RMS, i.e.
	\begin{equation}
	\gamma_{ij}=\frac{\langle r_{i} r_{j}\rangle}{\sigma^2+\sigma_{\rm r}^2}\,.
	\end{equation}
	$\sigma_{\rm r}$ is the RMS of the coloured noise. The summation of index $i$ and $j$ runs within the pulse duration, which includes $N_{\rm box}$ data points.
	The distribution of the $S$ with coloured noise is thus
	\begin{equation}
	f(S|N_{\rm box})=\frac{1}{\sqrt{2\pi}}\left(\frac{S}{\langle S \rangle}\right)^{-\frac{1}{2}}e^{-\frac{S}{2\langle S\rangle}}\,.
	\label{eq:sdis}
	\end{equation}
Note here that the distribution function $f(S|N_{\rm box})$ depends on the pulse 
width, since it depends on $N_{\rm box}$.
	The term $\sum_{ij}\gamma_{ij}$ can be simplified with the help of coloured noise power spectrum density (${\cal S}(f)$), which is a Fourier transform of the two point correlation function. By definition, we have
    \begin{equation}
	\gamma_{i,j}=\frac{1}{\sigma^2+\sigma_{\rm r}^2}\int_{0}^{\infty} {\cal S}(f) e^{2 \pi i f (t_i-t_j)} {\,\rm d}f\,,
	\end{equation}
and, after interchanging the order of summation and integration, one gets
	\begin{equation}
	\sum_{ij} \gamma_{i,j}=\frac{1}{\sigma^2+\sigma_{\rm r}^2}\int_{0}^{\infty}{\cal S}(f) \frac{\sin^2 (f N_{\rm box} \pi \Delta T)}{\sin^2(f\pi \Delta T)}{\,\rm d}f\,,
	\end{equation}
with $\Delta T$ being the sampling time.
The above equation can be further simplified, if we replace the discrete 
summation using the integral, i.e. assuming $\sum_i \simeq \frac{N}{ 
T}\int_{0}^{T} \, dt$, we have
	\begin{equation}
	\sum_{ij} \gamma_{i,j}=\frac{1}{\sigma^2+\sigma_{\rm r}^2}\int_{0}^{\infty}{\cal S}(f) \frac{\sin^2 (f N_{\rm box} \pi \Delta T)}{(f\pi \Delta T)^2}{\,\rm d}f\,,
	\end{equation}
There are two major types of correlated noise, 1) the red noise dominated by low 
frequency components and 2) quasi-monochromatic noise dominated by a single 
frequency component. 
	
For case 1), we can assume that the noise spectrum is a power-law function, i.e. 
$S(f)=A f^{-\alpha}$, where $\alpha$ is the spectral index. 
one gets (similar computation can be found in \citet{LBJ12}),
\begin{eqnarray}
	\sum_{i,j}&&\gamma_{i,j}=N_{\rm box}^2\frac{\sigma_{\rm 
	r^2}}{\sigma^2+\sigma_{\rm r}}\frac{\alpha-1}{\pi} \times \nonumber \\
	&&\left[ (2\pi)^{\alpha} \Gamma\left(-1-\alpha\right)\sin(\frac{\pi \alpha}{2}) +\right. \nonumber\\
	&&\left.\frac{2 \pi}{\alpha^2-1} \,_1F_2\left(\frac{1-\alpha}{2}; \frac{3}{2},\frac{3-\alpha}{2}, -\pi^2\right)\right]\,.
	\end{eqnarray}
Here $\Gamma$ is the gamma function, and $\,_1F_2$ is the hypergeometric 
function. For spectral index runs within $\alpha\in [1,10]$, the above result can be approximated by the 
average value
\begin{equation}
	\sum_{ij} \gamma_{i,j,{\rm power}}\simeq\frac{N_{\rm box}^2}{100}\frac{\sigma_{\rm 
	r}^2}{\sigma^2+\sigma_{\rm r}^2}\,,
	\end{equation}
which leads to  \EQ{eq:scale} with $\kappa\simeq 1/100$.
For the quasi-monochromatic case, the spectral density is approximated by the 
Dirac's $\delta$ function, and one has
	\begin{equation}
	\sum_{ij}\gamma_{i,j,{\rm mono}}=N_{\rm box}^2\frac{\sigma_{\rm r}^2}{\sigma^2+\sigma_{\rm r}^2}{\rm sinc}^2(\pi f_0 T)\,,
	\end{equation}
	with $f_0$ being the frequency of monochromatic noise. For $fT\ll1$, we have $\sum_{ij}\gamma_{i,j, {\rm mono}}=N_{\rm box}^2 \sigma_{\rm r}^2/(\sigma^2+\sigma_{\rm r}^2)$, i.e. $\kappa={\rm sinc}^2(\pi f_0 T)$.

For most of the FRB searches, one will only record the signal above certain 
threshold, saying $S\ge S_0$. In order to compare the observation, we compute 
here the expectation and standard deviation of $S$ with threshold selection 
$S\ge S_0$. From the distribution function, we have
\begin{eqnarray}
	\left.\langle S\rangle \right|_{S\ge S_0}&=&\frac{\int_{S_0}^{\infty} f(S) S 
	ds}{\int_{S_0}^{\infty} f(S) ds} \,,\\
	\left.\langle S^2\rangle \right|_{S\ge S_0}&=&\frac{\int_{S_0}^{\infty} f(S) S^2 ds}{{\int_{S_0}^{\infty} f(S) ds}}\,,
	\end{eqnarray}
which produce
	\begin{eqnarray}
	\left.\langle S\rangle \right|_{S\ge S_0}=\frac{\langle S\rangle \text{erfc}\left(\sqrt{\frac{S_0}{2\langle S\rangle}}\right)+\sqrt{\frac{2 S_0 \langle S\rangle}{\pi }} e^{-\frac{S_0}{2
				\langle S\rangle}}}{{\rm erfc}(\sqrt{\frac{S_0}{2 \langle S \rangle }})}\,, \\
	\left.\langle S^2\rangle \right|_{S\ge S_0}=\frac{3 \langle S\rangle^2 \text{erfc}\left(\sqrt{\frac{S_0}{2\langle S\rangle}}\right)+ \sqrt{\frac{2 S_0 \langle S\rangle}{\pi }}(S_0+3 \langle S\rangle ) e^{-\frac{S_0}{2
				\langle S\rangle}}}{{\rm erfc}(\sqrt{\frac{S_0}{2 \langle S \rangle }})}
	\end{eqnarray}
If one is interested in the distribution of $S$ for all pulses found in the 
given data (denoted as ${\cal F}(s)$). We can sum over the possible choice of 
$N_{\rm box}$. The number of 
independent pulse sample for the given data is $T/(\Delta T N_{\rm box})$, and the 
distribution of $S$ for all \emph{observed} pulses is given by
	\begin{equation}
	{\cal F}(s)\propto\sum_{N_{\rm box}}  f(S|N_{\rm box}) \frac{T}{\Delta T N_{\rm box}}
	\end{equation}

\bsp	
\label{lastpage}
\newpage
\section{\textsc{BEAR} detection plot}
\label{sec:bearplot}

\begin{figure*}
    \centering
    \includegraphics[width=\textwidth]{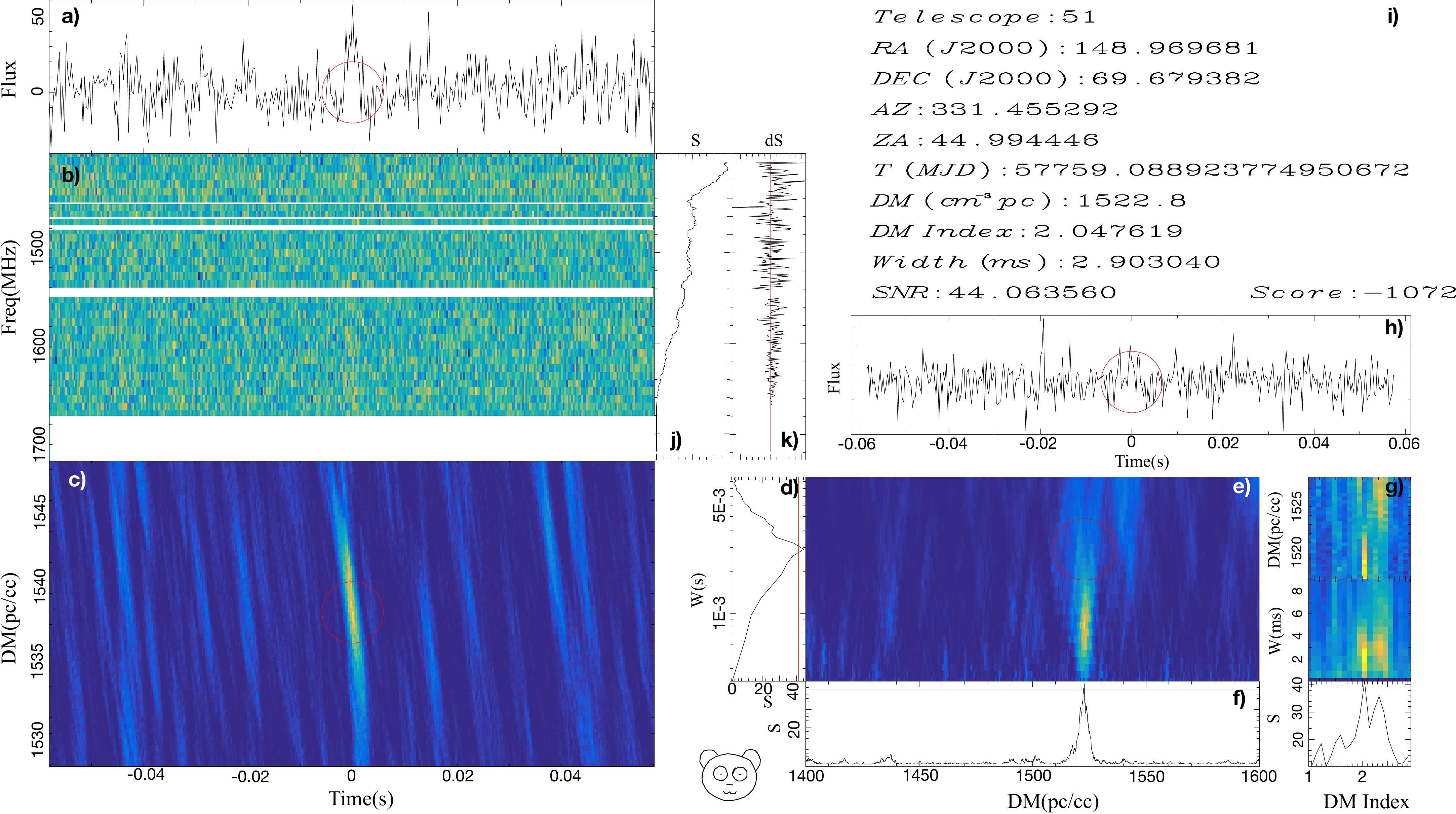}
\caption{A FRB candidate found along the M82 direction. The plot is produced by \textsc{BEAR}. a) the de-dispersed candidate pulse profile with a red circle indicating the pulse time of 
arrival, b) the dynamical spectrum after de-dispersion, c) the likelihood ratio test statistics $S$ as a function of DM and time, d) $S$ as a function of pulse width, e) $S$ as a function of DM 
(x-axis) and pulse width (y-axis), red circle label the best estimated DM 
value, f) $S$ as a function of DM, where red horizontal lines indicates false alarm rate of $10^{-10}$, 
g) DM, pulse width and $S$ as a function of DM index, h) the 
time series de-dispersed at 0 DM, i) the basic information of the 
pulse, j) the integration of $S$(x-axis) changes over frequency channel, k) the contribution of $S$ from each 
frequency channel. \label{fig:bear}}
\end{figure*}

In this section, we show the \textsc{BEAR} candidate sifting plot for the signal we found, i.e. \FIG{fig:bear}. As one can see from sub-panel j), the signal is contributed from channels across rather wide bandwidth. Sub-panel h) shows that the burst can not be found in the  zero-dm time series, which verify the dispersive nature of the burst, and sub-panel g) further indicates that the DM index is around -2.

\end{document}